\documentclass{article}
\usepackage{spconf,amsmath,amssymb,epsfig,caption}

\newcommand{\bom}[1]{\boldsymbol{#1}}
\newcommand{\bo}[1]{\mathbf{#1}}
\usepackage{subfigure}
\usepackage{hhline}
\usepackage[ruled,boxed,boxruled]{algorithm2e}
\graphicspath{{./}}

\newcommand{\s}{\bo x}  
\newcommand{\es}{x}      

\renewcommand{\a}{\bo a}  
\newcommand{\y}{\bo y} 
\newcommand{\x}{\bo x} 
 
\newcommand{\g}{\bo g}

\newcommand{\e}{\bo e}  
\newcommand{\ee}{e}      
\newcommand{\veps}{\bom \varepsilon}

\newcommand{\A}{\bo A} 
\newcommand{\V}{\bo V} 
\newcommand{\W}{\bo W} 
\newcommand{\I}{\bo I}

\newcommand{\Om} {\Gamma} 
 
\newcommand{\Gam}{\Gamma} 
\newcommand{\al}{\alpha}
\newcommand{\be}{\beta}
\newcommand{\sig}{\sigma}
\newcommand{\hsig}{\hat \sigma}

\newcommand{\supp}{\mathrm{supp}} 
\newcommand{\diag}{\mathrm{diag}}
\newcommand{\eps}{\varepsilon} 
\newcommand{\Med}{\mathrm{Med}}

\newcommand{\beq}{\begin{equation}}
\newcommand{\eeq}{\end{equation}}
\newcommand{\bmat}{\begin{pmatrix}}
\newcommand{\emat}{\end{pmatrix}}
\newcommand{\beqa}{\begin{eqnarray}}
\newcommand{\eeqa}{\end{eqnarray}}

\newcommand{\R}{\mathbb{R}}

\newcommand{\E}{\mathbb{E}}

\newcommand{\mm}{\bom \Phi}   
\newcommand{\mb}{\bom \phi}    
\newcommand{\ndim}{M}             
\newcommand{\pdim}{N}             
\newcommand{\kdim}{K}             

\newcommand{\paino}{\it}
\newcommand{\MSE}{\mathrm{MSE}}
\newcommand{\SNR}{\mathrm{SNR}}
\newcommand{\MAD}{\sigma_{\tiny{\mbox{MAD}}}}
\newcommand{\SD}{\sigma_{\tiny{\mbox{SD}}}}
\newcommand{\MEAD}{\sigma_{\tiny{\mbox{MeAD}}}}

\def\x{{\mathbf x}}

\title{Robust iterative hard thresholding for compressed sensing}
\name{Esa Ollila, Hyon-Jung Kim and Visa Koivunen}
\address{Aalto University\\
Dept. of Signal Processing and Acoustics \\
P.O.Box 13000, FI-00076 Aalto, Finland}

\begin{document}
%
\maketitle
\begin{abstract}
Compressed sensing (CS) or sparse signal reconstruction  (SSR)   is a signal processing technique that exploits the fact that acquired data 
can have a  sparse representation in some basis.
 One popular technique to reconstruct or approximate the unknown sparse signal  is the iterative hard thresholding (IHT) 
 which however performs very poorly under non-Gaussian noise conditions 
or in the face of outliers (gross errors). In this paper, we propose a robust IHT method based on ideas from $M$-estimation that estimates the sparse signal and the scale  of the error distribution simultaneously.
The method has a negligible performance loss compared to IHT under Gaussian noise, but superior performance under heavy-tailed non-Gaussian noise conditions. \end{abstract}
\begin{keywords}
Compressed sensing, iterative hard thresholding, $M$-estimation, robust estimation
\end{keywords}
\section{Introduction} \label{sec:intro}

The compressed sensing (CS) 
problem can be formulated as follows \cite{sparse_book:2010}. 
 Let $\y=(y_1,\ldots,y_\ndim)^\top$ denote the observed data ({\paino measurements}) modelled as 
\begin{align} \label{sampling_eq} 
 \y = \mm \s + \veps  
\end{align} 
where $\mm 
= \bmat \mb_1 & \cdots & \mb_\ndim\emat^\top$ is $\ndim \times \pdim$ {\paino measurement matrix}  
with more column vectors 
than row vectors $\mb_i$ (i.e., $\pdim>\ndim$),  $\s=(\es_1,\ldots,\es_\pdim)^\top$ is the unobserved  {\paino signal vector}  and 
$\veps = (\eps_1,\ldots,\eps_\ndim)^\top$ is the (unobserved) {\paino random noise} vector.
It is assumed that the signal vector $\s$ is {\paino $\kdim$-sparse}
(i.e., it has $\kdim$ {\it non-zero} elements) or is {\paino compressible} (i.e., it has a representation
whose entries decay rapidly when sorted in a decreasing order)
The signal {\paino support} (i.e., the locations of non-zero  elements) is denoted as $\Gam = \supp(\s) = \{   j \ : \  \es_j  \neq 0 \}$.  	
Then, we aim to reconstruct or approximate the
signal vector $\s$ by $\kdim$-sparse representation knowing only the acquired vector $\y$, the measurement matrix $\mm$ and
the sparsity $\kdim$. 

A $\kdim$-sparse estimate of $\s$ can be found by solving the optimization 
$\min_{ \s } \| \y -\mm \s \|_2^2$ subject to   $\| \s \|_0 \leq \kdim$, 
where  $\| \cdot \|_0$ denotes the $\ell_0$ pseudo-norm, $\|\s\|_0 =  \# \{   j \ : \  \es_j \neq 0 \}$ . 
This optimization problem is known to be NP-hard and hence suboptimal approaches have been under active research;  see \cite{sparse_book:2010} for a review. The widely used methods developed for estimating $\s$ such as 
{\paino Iterative Hard Thresholding (IHT)} \cite{blumensath_davies:2009,blumensath_davies:2010} 
are shown to perform very well provided that suitable conditions (e.g., restricted isometry property on $\mm$ and  non impulsive noise conditions) are met.  
Since the recovery bounds of IHT depend linearly on $\| \veps \|_2$, the method often fails to provide accurate reconstruction/approximation  under the heavy-tailed or spiky non-Gaussian noise.  

 Despite  the vast interest in CS/SSR  during the past decade, {\it sparse and robust signal reconstruction  methods} that  are resistant to heavy-tailed non-Gaussian noises or outliers  have  appeared in the literature only recently; e.g, \cite{carrillo_etal:2011,parades_arce:2011,alireza_etal:2012}. 
In \cite{alireza_etal:2012} we proposed a robust IHT method using a robust loss function with a preliminary estimate of the scale, called the generalized IHT. The Lorentizian IHT (LIHT) proposed in \cite{carrillo_etal:2011} is a special case 
of our method using the Cauchy loss function. A major disadvantage of these methods is that they require a 
preliminary  (auxiliary) robust estimate of the scale parameter $\sigma$ of the error distribution. 
In this paper, we propose a novel IHT method that estimates  $\s$  and   $\sig$ simultaneously.

The  paper is organized as follows. Section~\ref{sec:robust} provides a review of the robust  $M$-estimation approach to regression using different robust loss/objective functions.  
We apply these approaches to obtain (constrained) sparse and robust estimates of $\s$ in the CS system model using the IHT technique. 
Section~\ref{sec:HubIHT}  describes the new robust IHT method and 
Section~\ref{sec:simul} provides extensive simulation studies  illustrating  the effectiveness of the method in reconstructing a $\kdim$-sparse signal in various noise conditions and 
SNR regimes.  

{\bf Notations:} 
For a vector $\a \in \R^{m}$, a matrix $\A \in \R^{n \times m}$ 
and an index set $\Gam =(\gamma_1,\ldots,\gamma_p)$ with $p<m$, $a_i$ denotes the $i$th component of $\a$, $\a_{i}$ denotes the $i$th column vector of $\A$ and 
 $\a_\Gam$  denotes the $p$-vector of $\a$
with elements $a_{\gamma_i}$ selected according to the support set $\Gam$.  
Similarly $\A_\Gam=(\a_{\gamma_1} \ \cdots \ \a_{\gamma_p})$ is an $n \times p$ matrix whose columns are selected from the columns of $\A$ according to the index set $\Gam$.


\section{Robust regression and  loss functions} \label{sec:robust}

We assume that the noise terms $\eps_i$ are  independent and identically distributed (i.i.d.) random variables from a continuous symmetric distribution 
and let $\sigma>0$ be the {\paino scale parameter} of the error distribution. The density of  $\eps_i$ is $f_\eps(e)= (1/\sig) f_0(e/\sig)$, where 
$f_0(\cdot)$ is the standard form of the density, e.g., $f_0(e)=(1/\sqrt{2\pi}) \exp(- \frac 1 2 e^2)$ in case of normal (Gaussian) error distribution. 
Let {\paino residuals} for a given (candidate) signal vector $\s$ be   
$
\ee_i \equiv e_i(\s) =  y_i - \mb^\top_i  \s 
$
and write $\bo e  \equiv \e(\s)=(\ee_1,\ldots,\ee_\ndim)^\top=\y- \mm \x$ for the vector residuals. 
When  $\ndim > \pdim$ and no sparse approximation  is assumed for $\s$ (i.e., unconstrained overdetermined problem), \eqref{sampling_eq} is just a conventional {\paino regression model}. We start with a brief review of the robust $M$-estimation method.

A common approach to obtain a robust estimator of regression parameters 
is to replace the  least squares (LS) or $\ell_2$-loss function $\rho(e)= \frac 1 2 e^2$
by a {\paino robust loss function} which downweights large residuals. 
Suppose that we have obtained an (preliminary, a priori) estimate $\hat \sigma$ of the scale parameter $\sigma$.  
Then, a robust $M$-estimator $\hat \s$ of $\s$ can be obtained by solving the optimization problem 
\beq \label{eq:opt}
\hat \s = \arg \min_{\s}  \sum_{i=1}^\ndim \rho \bigg( \frac{y_i - \mb^\top_i  \s}{\hat \sigma}  \bigg) 
\eeq 
where $\rho$ is a continuous, even function increasing in $\ee \geq 0$.   The {\paino $M$-estimating equation} approach 
is to find  $\hat \s$ that solves  
$\sum_{i=1}^\ndim \psi\big( (y_i - \mb^\top_i  \s)/\hsig \big) \mb_{i} = \bo 0$,  
where $\psi$  is a continuous and odd function ($\psi(-e)=-\psi(e)$), referred to as a {\paino score function}.  
When $\psi = \rho'$,  a stationary point of the objective function in \eqref{eq:opt} is a solution to  the estimating equation. 

A commonly used loss function  is {\bf Huber's loss function} 
which  combines $\ell_2$  
and $\ell_1$ loss functions and is defined as  
\beq \label{eq:hubrho} 
\rho_{\rm H}(e) =  
\begin{cases} \frac{1}{2} e^2, &\mbox{for  $|e| \leq c$} \\   c |e| - \frac{1}{2} c^2, &\mbox{for  $|e| > c$}, \end{cases}
\eeq 
where $c$ is a  user-defined {\paino tuning constant} that influences the degree of robustness 
and efficiency of the method. The following choices, 
$c_1=1.345$ and $c_2 =  0.732$,  
yield 95 and 85 percent (asymptotic) relative efficiency compared to LSE of regression  in case of Gaussian errors. 
Huber's loss function is differentiable and convex function, and 
the score function $\psi=\rho'$ is a {\paino winsorizing (clipping, trimming) function}   
\begin{align*}
 \psi_{\rm H}(e) &=
  \max[-c,\min(c,e)] 
  = \begin{cases} e, &\mbox{for  $|e| \leq c$} \\  c \, \mathrm{sign}(e), &\mbox{for $e>c$}  
  \end{cases} 
\end{align*} 
The smaller the $c$, the more downweighting (clipping) is done to the residuals. 

\section{Robust IHT } 
\label{sec:HubIHT}


The  problem of generalized  IHT of \cite{alireza_etal:2012} is in how to  obtain an accurate and robust preliminary scale estimate $\hsig$. 
To circumvent the above problem, we propose to estimate the $\s$ and $\sig$ simultaneously (jointly).  
To do this elegantly, we propose to minimize  
\begin{align}
&   Q(\s,\sig) = \sig \sum_{i=1}^\ndim  \rho \bigg(  \frac{y_i - \mb_i^\top \s}{\sig} \bigg)  +  (\ndim-\kdim) \al   \sig   \label{eq:Qfun} \\
&\mbox{ subject to }  \quad \| \s \|_0 \leq \kdim ,   \notag
\end{align}
where $\rho$ is a convex loss function which should verify
$
\lim_{|x| \to \infty}  \rho(x)/|x|  = c \leq \infty
$
and $\al>0$ is a scaling factor  chosen so that the solution $\hsig$ 
is Fisher-consistent for $\sigma$ when  $\eps_i \sim \mathcal N(0,\sig^2)$. This is achieved by setting  $\al =  \E[\chi(u)]$,  
where $u\sim \mathcal N(0,1)$ 
and 
$\chi(e)=\psi(e)e - \rho(e)$. 
Note that a multiplier $(\ndim-\kdim)$ is used in the second term of \eqref{eq:Qfun} instead of $\ndim$ in order to reduce the bias of the obtained scale estimate $\hsig$ at  small sample lengths.  
The objective function $Q$ in \eqref{eq:Qfun} was proposed for  joint estimation of regression and scale 
by Huber (1973)  \cite{huber:1973} and is often referred to as  {\paino "Huber's proposal 2"}. 
 Note that $Q(\s,\sig)$ is a convex function of $(\s,\sig)$ 
 which allows to derive a simple convergence proof of an iterative algorithm to compute the solution $(\hat \s,\hsig)$.

Let us choose Huber's loss function $\rho_{\rm H}(e)$ in eq. \eqref{eq:hubrho} as our choice  of $\rho$ function. 
In this case $\chi$-function  
becomes $\chi_{\rm H}(e)= \frac{1}{2} \psi_{\rm H}^2(e)$ and 
the scaling factor $\al =\be/2$ can be computed as 
\beq \label{eq:beta}
\beta = 2 \{ c^2(1- F_{G}(c))  + F_{G}(c) - 1/2 - c \, f_{G}(c) \},
\eeq 
where $F_{G}$ and $f_{G}$ denote the c.d.f and the p.d.f. 
of $\mathcal N(0,1)$ distribution, respectively, and 
 $c$ is the downweighting threshold  of Huber's loss function. 
The algorithm for finding the solution to \eqref{eq:Qfun} is given in Algorithm~\ref{alg:HubIHT}. Therein   
 $H_\kdim(\cdot)$ (in Step 5 and initialization step) denotes the hard thresholding operator that sets all but the largest (in magnitude) $\kdim$ elements
of its vector-valued argument to zero.  

\begin{algorithm}[!t]
\caption{Huber IHT (HIHT) algorithm} \label{alg:HubIHT}
{\bf Input}:  $\y$,  $\mm$, sparsity $\kdim$ and trimming threshold $c$. \\ 
 {\bf Output}:   $( \s^{n+1}, \sig^{n+1}, \Gam^{n+1})$  estimates of $\s$, $\sig$ and $\Gam$.  \\
 {\bf Initialization}: Set $\s^0=\bo 0$, $\sigma^0 = 1$. 
 Compute the scaling factor $\beta=\beta(c)$ 
and the initial signal support $\Gam^0 =\supp\big( H_\kdim(\mm^\top \y_\psi) \big)$, where $\y_\psi= \psi_{\rm H}(\y)$.   
\begin{enumerate}
\item[{\bf For }]  $n=0,1, \ldots,$ iterate  the steps 

\item {\it Compute the residuals} 
$\e^{n} = \y-\mm \x^{n}$ 

\item {\it Update the value of the scale}:  
\[
(\sig^{n+1})^2 = \frac{(\sigma^n)^2}{ (\ndim - \kdim)\beta} \sum_{i=1}^\ndim \psi_{\rm H}^2\bigg( \frac{\ee_i^{n}}{\sigma^n}  \bigg)
\]

\item {\it Compute the pseudo-residual and  the gradient update}
\[
\e_\psi^{n} = \psi_{\rm H} \bigg( \frac{\e^n}{\sig^{n+1}} \bigg) \sig^{n+1}  \  \mbox{and} \ \, \g  = \mm^\top \e_\psi^n
\] 

\item  {\it Compute the stepsize $\mu^n$}  using  \eqref{eq:mu0}  if $n=0$ and \eqref{eq:mun} otherwise 

\item  {\it Update the value of the signal vector and the support} 
\[
\s^{n+1} = H_\kdim (\s^{n} + \mu^n \g ) \ \, \mbox{and} \ \,
\Gam^{n+1} = \supp(\s^{n+1}) 
\]

\item {\it Approve the updates}  $(\s^{n+1}, \Gam^n)$   {\it or recompute  them} (discussed later).

 \item[{\bf Until}] 
 $
 \frac{\| \s^{n+1} - \s^n\|^2}{\|\s^{n}\|^2} < \delta, 
 $
 where $\delta$ is a predetermined tolerance/accuracy level (e.g., $\delta=1.0^{-6}$). 
\end{enumerate} 
\vspace{-0.2cm}
\end{algorithm}



{\bf Computing the stepsize $\mu^n$ in Step 4.}  
Assuming we have identified the correct signal support at $n$th iteration, an optimal step size 
can be found in gradient ascent direction $\s_{\Gam^n} + \mu^n \g_{\Gam^n}$   by  
solving 
\beq
\mu^n_{opt} = \arg \min_\mu     \sum_{i=1}^\ndim \rho\bigg( \frac{y_i-  [\mb_i]_{\Gam^n}^\top (\s^n_{\Gam^n}+\mu \g_{\Gam^n}) }{\sig^{n+1}}   \bigg). 
\eeq 
Since closed-form solution can not be found for $\mu^n_{opt}$, we aim at finding a good approximation in closed-form. By writing 
$v(e)=\rho(e)/e^2$, we can express the problem as 
\beq
\mu^n_{opt} = \arg \min_\mu     \sum_{i=1}^\ndim  v_i^n(\mu)   \big(y_i-  [\mb_i]_{\Gam^n}^\top (\s^n_{\Gam^n}+\mu \g_{\Gam^n} ) \big)^2 
\eeq 
where 
$v_i^n(\mu) =  v\big(  (y_i-  [\mb_i]_{\Gam^n}^\top (\s^n_{\Gam^n}+\mu \g_{\Gam^n}))/\sig^{n+1} \big)$ 
depend on $\mu$. If we replace $v_i^n(\mu)$ 
by its approximation $v_i^n=v_i^n(0)$,  we can find stepsize (i.e., an approximation of $\mu^n_{opt}$) in closed-form. 
Hence, when the iteration starts at $n=0$, we calculate the stepsize $\mu^0$ in Step 4 as 
\beq \label{eq:mu0}
\mu^0 = \frac{  (\e^0)^\top \V^0 \mm_{\Gam^0} \g_{\Gam^0}}{ \g^\top_{\Gam^0} \mm_{\Gam^0}^\top \V^0 \mm_{\Gam^0} \g_{\Gam^0} }, 
\eeq 
where $\V^0=\diag(v_1^0,\ldots,v_\ndim^0)$. When iteration proceeds (for $n=1,2,\ldots$), the current support $\Gam^n$ and the signal update $\s^n$ are more accurate estimates of $\Om$ 
and  $\x$.  Hence, when $n \geq 1$,  
we find an approximation of 
$\mu_{opt}^n$ by solving  
\beq \label{eq:muapprox}
\mu^n = \arg \min_{\mu} \sum_{i=1}^n w_i^n  \Big( y_i-  [\mb_i]_{\Gam^n}^\top (\s^n_{\Gam^n}+\mu \g_{\Gam^n})  \Big)^2 
\eeq 
where the "weights" $w_i^n$ are defined as 
$w_i^n = w_{\rm H}\big(  (y_i-  [\mb_i]_{\Gam^n}^\top \s^n_{\Gam^n}) /\sig^{n+1} \big)$ with  $w_{\rm H}(e)=\psi_{\rm H}(e)/e$ being the Huber's weight function. 
The solution to \eqref{eq:muapprox} is 
\beq \label{eq:mun}
\mu^{n} =  \frac{\g_{\Gam^n}^\top \g_{\Gam^n}}{  \g_{\Gam^n}^\top \mm_{\Gam^n}^\top \W^n  \mm_{\Gam^n}\g_{\Gam^n} }, 
\eeq
where $\W^n = \diag(w_1^n,\ldots,w_\ndim^n)$. 

 {\bf Approving or recomputing the updates $(\s^{n+1},\Gam^{n+1})$ in Step 6.}  
 We accept the updates if 
$Q(\x^{n+1},\sig^{n+1}) < Q(\x^n,\sig^n)$,   otherwise
we  set $\mu^n  \leftarrow \mu^n/2$ and go back to Step 5 and recompute new updates.

{\bf Relation to IHT algorithm.}  
Consider the case that trimming threshold $c$ is arbitrarily large ($c \to \infty$). Then it is easy to show that the proposed Huber IHT method coincides with IHT \cite{blumensath_davies:2009,blumensath_davies:2010}.  This follows as Step 2 can be discarded as it does not have any effect on Step~3 because $\e_\psi^n=\e^n$ for very large $c$ (as $\psi_{\rm H}(e)=e$). 
Furthermore, now $\V^0 = \I$ and $\W^n=\I$, so  the optimal stepsizes \eqref{eq:mu0} and \eqref{eq:mun} reduce to the one used in the normalized IHT algorithm 
\cite{blumensath_davies:2010}.

\section{Simulation studies} \label{sec:simul}

{\bf Description of the setup and performance measures}. 
The elements of the measurement matrix $\mm$ are drawn from $\mathcal{N}(0,1)$ distribution  after which the columns of are normalized to have unit norm.  
The $\kdim$ nonzero coefficients of $\s$ are set to have   equal amplitude $\sigma_s = |\es_i|=10$ for all $i \in \Om$, equiprobable signs 
and 
$\Om=\supp(\s)$ is randomly chosen from $\{1,\ldots,\pdim\}$  without replacement 
for each trial. 
The {\paino signal to noise ratio (SNR)}  is 
$
\SNR(\sig) = 20 \log_{10} (\sigma_s/\sig)  $
and depends on the scale parameter $\sigma$ of the error distribution. In case of Gaussian errors, the scale equals the {\paino standard deviation} (SD)  $\SD= \sqrt{\E[|\eps|^2]}$,   
in case of Laplacian errors, the {\paino mean absolute deviation} (MeAD) $\MEAD = \E[|  \eps |]$,  
and in case of Student's $t_\nu$-distribution with  degrees of freedom (d.o.f.) $\nu \geq 0$,  the   {\paino median absolute deviation} (MAD) $\MAD= \Med(|\eps|)$. Note that 
SD does not exist for $t_\nu$-distribution with $\nu\leq 2$.  As performance measures 
we use the {\paino mean squared error}
$
\MSE(\hat \s) = \frac{1}{Q} \sum_{q=1}^Q {\| \hat \s^{[q]} - \s^{[q]} \|_2^2}
$
and the {\paino probability of exact recovery}  
\[
\mbox{PER}(\hat \s)  = \frac{1}{Q}\sum_{q=1}^Q  \mathrm{I}(\hat \Om^{[q]} = \Om^{[q]}),
\]
where $\mathrm{I}(\cdot)$ denotes the indicator function, 
$\hat{\s}^{[q]}$ and $\hat \Om^{[q]} =\supp(\hat \s^{[q]})$ denote the  estimate of the $K$-sparse signal $\s^{[q]}$   and the signal support $\Om^{[q]}$  for the $q$th trial,  respectively. 
The number of Monte-Carlo trials is $Q=2000$,  $\ndim=512$,  $\pdim=256$ and the sparsity level is $\kdim=8$. 
The methods included in the study are  {\bf IHT} (referring to the normalized IHT method \cite{blumensath_davies:2010}), 
 {\bf LIHT}  (referring to LIHT method of  \cite{carrillo_etal:2011}) and 
 {\bf HIHT-$c_i$}, $i \in \{1,2\}$  (referring to the Huber IHT method of Algorithm~1 using trimming thresholds $c_1=1.345$ and $c_2 =  0.732$). 

\begin{table}[!t]
    \caption{PER rates of the methods under $t_\nu(0,\MAD)$ distributed noise at different $\SNR(\MAD)$ and d.o.f.  $\nu$. } \label{tab:tdist}
    \vspace{-0.2cm}
{\small
    \begin{tabular}{| c  | c | c | c | c | c | c | c | c | }
        \hline 
        &  \multicolumn{8}{c|}{Degrees of freedom $\nu$ } \\ 
        \hhline{~--------}
 Method                            &  1     &  1.25  & 1.5   & 1.75 &  2       &  3    &    4     & 5 \\  \hline 
                \multicolumn{9}{|c|}{$\SNR(\MAD)=$  40 dB} \\ \hline
                            IHT                   &  .51  &   .86   &  .95  &  .99  &  .99  &  1.0  &   1.0   &  1.0\\ 
               LIHT                  & 1.0   &   1.0   &  1.0  &  1.0  &   1.0   &  1.0  & 1.0  &   1.0  \\ 
               HIHT-$c_1$ & 1.0   &   1.0   &  1.0  &  1.0  &   1.0   &  1.0  & 1.0  &   1.0  \\  
              HIHT-$c_2$  & 1.0   &   1.0   &  1.0  &  1.0  &   1.0   &  1.0  & 1.0  &   1.0  \\  
           \hline \hline
                \multicolumn{9}{|c|}{$\SNR(\MAD)=$  20 dB} \\ \hline
             IHT                     &   0   &       0  &    0   &  .01 &  .04    &  .31  &  .57  &   .72 \\    
              LIHT                   & .09  &   .10   &  .13  & .13  &  .17    &  .22  & .24   &   .24  \\
              HIHT-$c_1$  & .46  &   .61   & .70    & .77  &  .81   &  .90  &  .92   &  .93  \\        
              HIHT-$c_2$  &  .61 &  .66    &  .72   & .73  &  .74   & .80   & .82   & .82 \\     
               \hline
       \end{tabular}}
\end{table}

\begin{figure}[!t]
\vspace{-0.2cm}
\centerline{(a)
\includegraphics[width=0.37\textwidth]{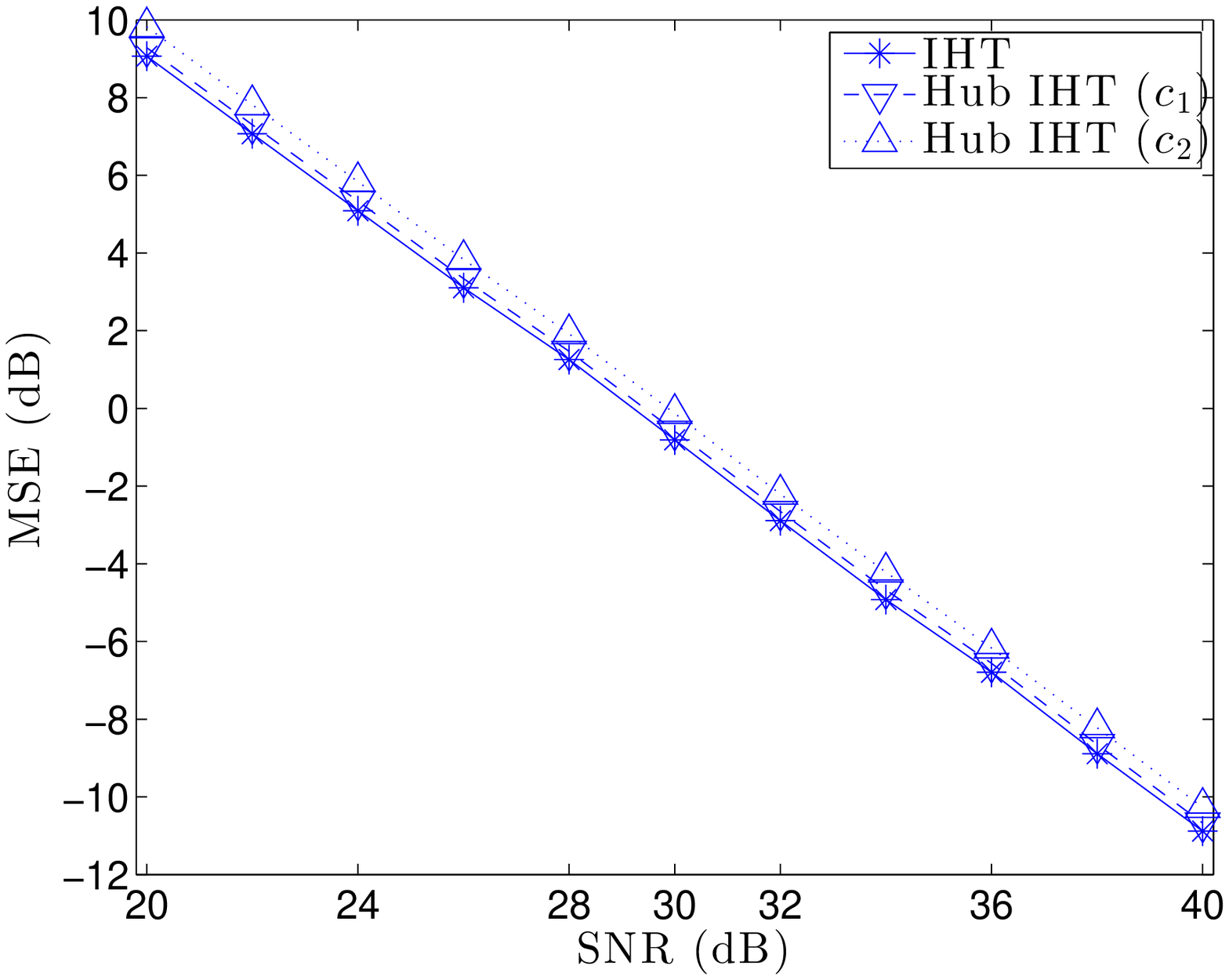}} 
\vspace{-0.4cm}
\centerline{(b)
\includegraphics[width=0.37\textwidth]{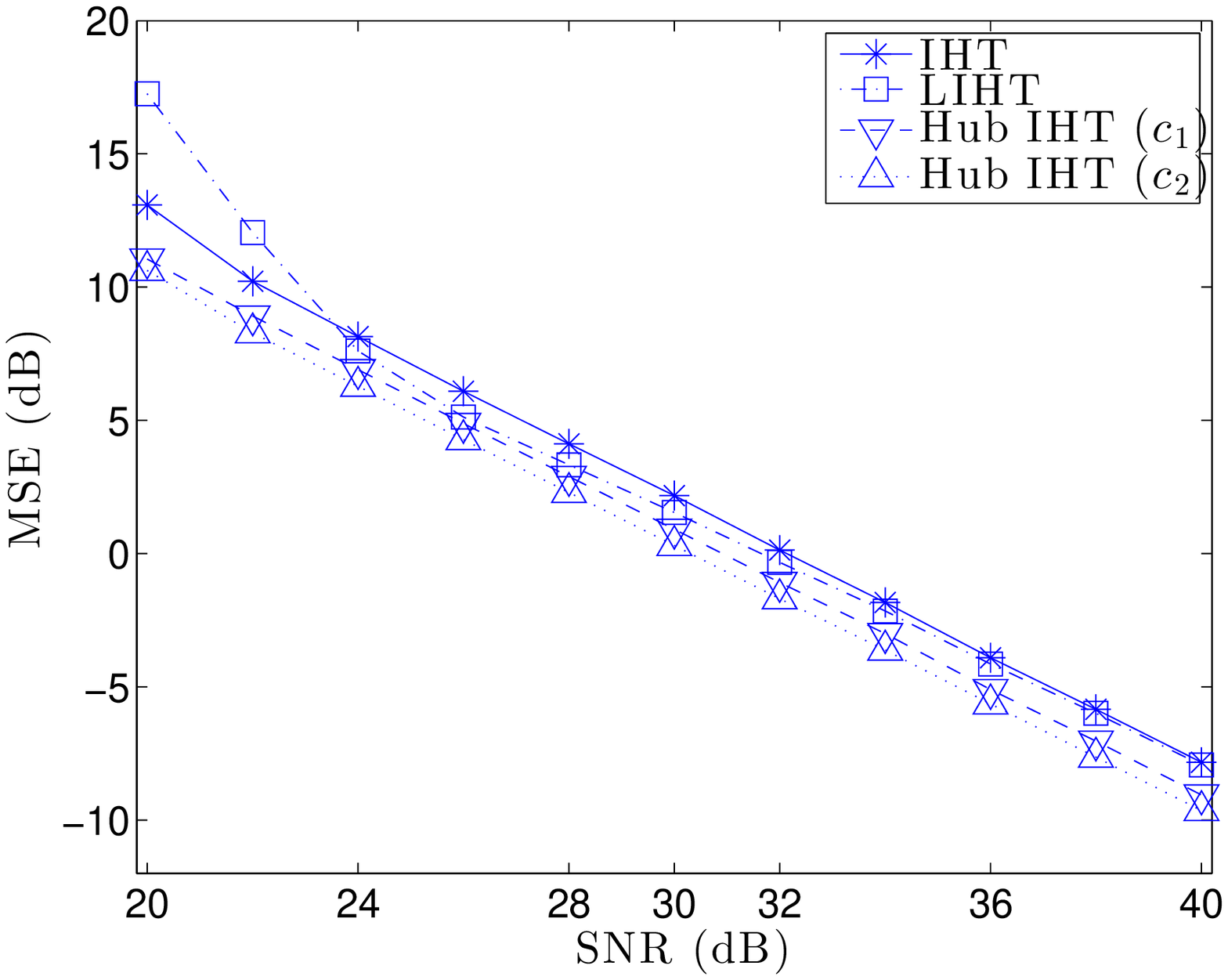}} 
\vspace{-0.2cm}
\caption{Average MSE of the methods as a function of $\SNR(\sig)$ 
under (a) $\mathcal N(0,\SD^2)$ noise and  (b) $Lap(0,\MEAD)$ noise.}   \label{fig:lap_gau} 
\end{figure}

{\bf Experiment I: Gaussian and Laplacian noise}: Figure~\ref{fig:lap_gau} depict the MSE as a function of SNR$(\sig)$  
in the Gaussian $\mathcal N(0,\SD^2)$ and  Laplace $Lap(0,\MEAD)$ noise distribution cases, respectively.   
In the Gaussian case,  the IHT has the best performance  but HIHT-$c_1$ suffers only a negligible 0.2 dB performance loss. 
LIHT  experienced convergence problems in the Gaussian errors simulation setup and hence was left out from this study.  These problems may be due to the choice of the preliminary scale estimate $\hsig$ used in LIHT 
which seems appropriate only for heavy-tailed distributions at high SNR regimes. 
In the case of Laplacian errors, the HIHT-$c_2$ has the best performance, next comes HIHT-$c_1$,  whereas LIHT has only slightly better performance compared to IHT in the high SNR regime [30, 40] dB.  
The performance loss of IHT as compared to HIHT-$c_2$ is 1.9 dB in average in  SNR [22 40], but jumps to 2.5 dB at SNR 20 dB. 
Note also that LIHT has the worst performance at low SNR regime. 
Huber IHT methods and the IHT had a full PER rate ($=1$) for all SNR in case of Gaussian errors. 
In case of Laplacian errors,  Huber IHT methods had again the best performance and they attained full PER rate at all SNR levels considered. The PER rates of LIHT decayed to 0.9900,  0.9300, and 0.7000  ad SNR = 24, 22, and 20 dB, respectively. The PER rate of IHT was full until SNR 20 dB at which it decayed to 0.97.

\begin{figure}[!t]
\centerline{
\subfigure[$\SNR(\MAD)= 40$ dB  \label{fig:tdista}]{\includegraphics[width=0.27\textwidth]{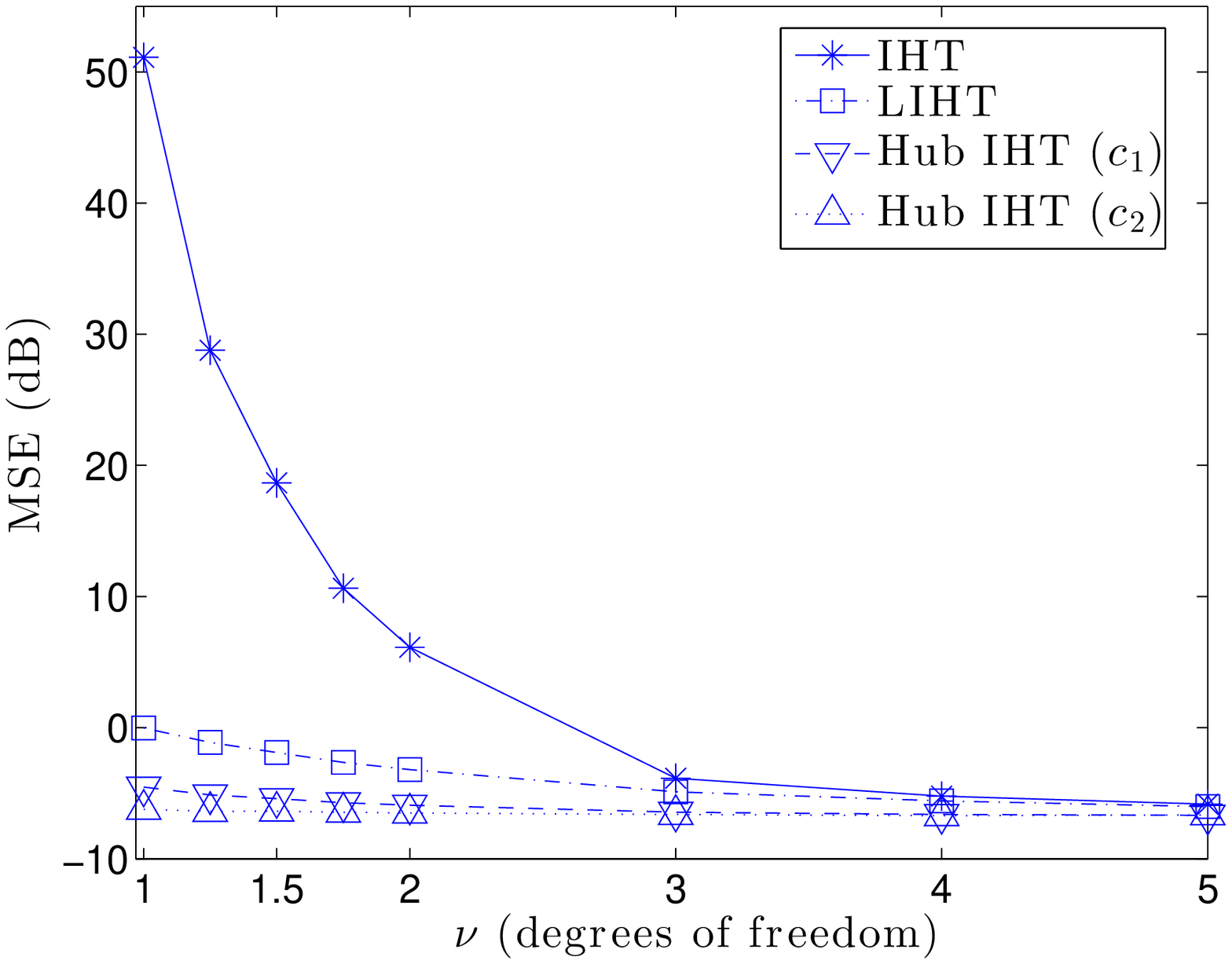}} \hspace{-0.7cm} \subfigure[$\SNR(\MAD)= 20$ dB  \label{fig:tdista3}]{
\includegraphics[width=0.27\textwidth]{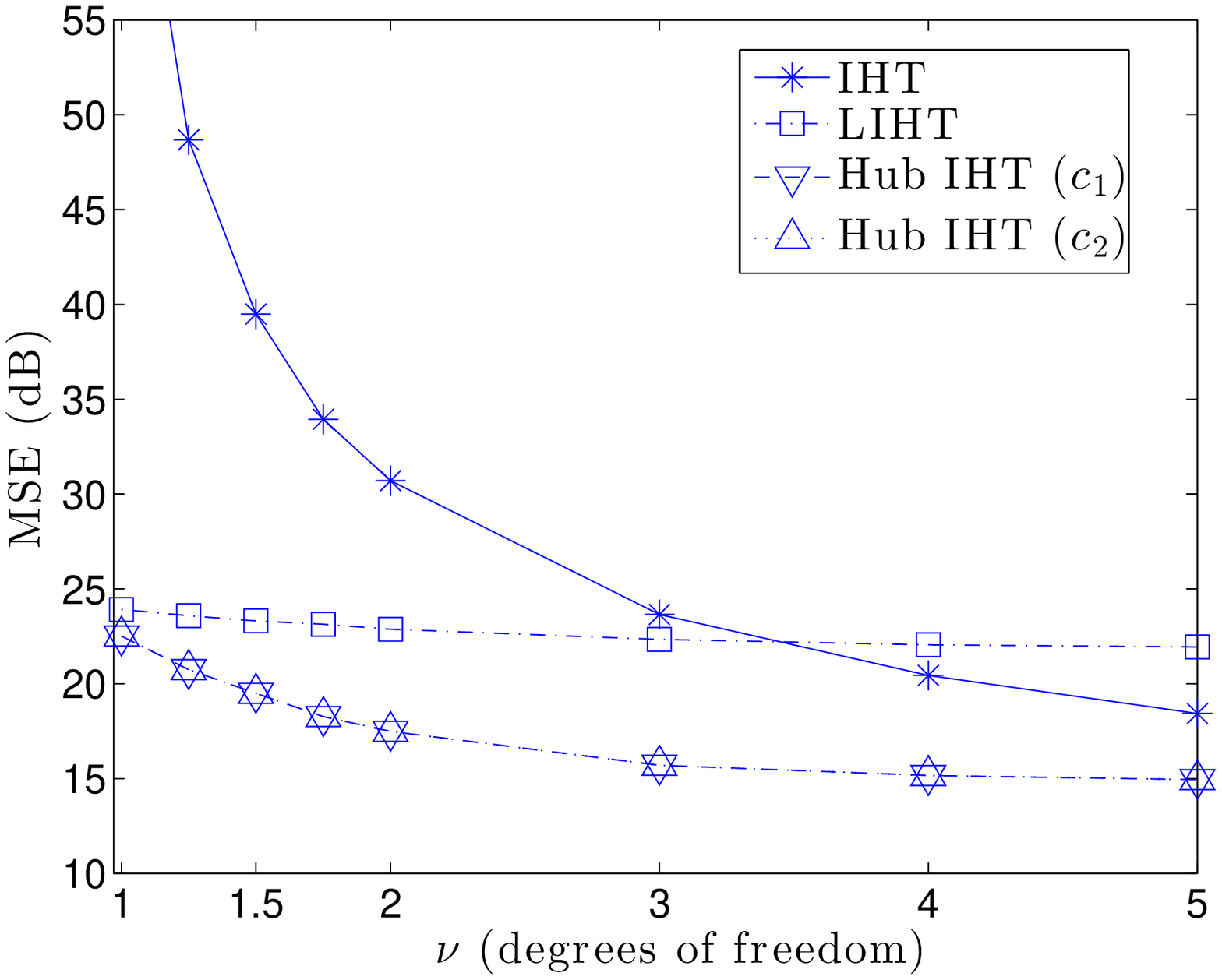}}}
\vspace{-0.4cm}
\caption{Average MSE  
of the methods 
under $t_\nu(0,\MAD)$ distributed  
noise  as a function of d.o.f. $\nu$.} 
\end{figure}

{\bf Experiment II: Student's $t_\nu$-noise}:  The noise terms are from Student's $t_\nu$-distribution, $t_\nu(0,\MAD)$. 
Figure~\ref{fig:tdista} and \ref{fig:tdista3} depict  
the  MSE as a function of d.o.f. $\nu$  at high (40 dB) and low (20 dB) $\SNR(\MAD)$ levels.  
Huber's IHT methods are outperforming the competing methods  in all cases. 
At high SNR 40dB in Figure~\ref{fig:tdista}, 
the Huber IHT  with $c_2$ is able to retain a steady MSE around -6.5 dB for all $\nu \in [1,5]$. The Huber IHT using $c_1$ is (as expected) less robust 
with slightly worse performance, but 
IHT is already performing poorly at $\nu=5$ and its performance  deteriorates 
at a rapid rate with decreasing $\nu$. The performance decay of  LIHT is much milder than that of IHT, yet it also has a rapid decay when compared to Huber IHT methods. 
The PER rates given in Table~\ref{tab:tdist} illustrate the remarkable performance of Huber's IHT methods which are able to maintain full recovery rates even at Cauchy distribution (when $\nu=1$) for SNR 40 dB. 
At low SNR 20 dB, only the proposed Huber IHT methods are able to maintain  good PER rates, whereas the IHT and LIHT 
provide estimates that are  completely corrupted.   

{\small 
\bibliographystyle{IEEEbib}
\bibliography{../../IEEE/bibtex/IEEEabrv,../../mybib/STATabrv,../../mybib/STATbib,../../mybib/ENGbib}
} 
\end{document}